\documentclass[a4paper,twoside]{article}

\usepackage{url}
\usepackage{calc}
\usepackage{amssymb}
\usepackage{amstext}
\usepackage{amsmath}
\usepackage{pslatex}
\usepackage{apalike}
\usepackage{algorithm2e}
\usepackage[bottom]{footmisc}
\usepackage{SCITEPRESS} 
\usepackage{graphicx}
\usepackage{booktabs}
\usepackage{mdframed}
\begin{document}

\title{Privacy-Aware Single-Nucleotide Polymorphisms (SNPs) using Bilinear Group Accumulators in Batch Mode}

\author{\authorname{William J Buchanan\sup{1}
        , Sam Grierson\sup{1} and Daniel Uribe\sup{2}}
\affiliation{\sup{1}Blockpass ID Lab, Edinburgh Napier University, Edinburgh, UK}
\affiliation{\sup{2}GenoVerse.io}
\email{\{b.buchanan, s.grierson2\}@napier.ac.uk}, 
}

\abstract{
Biometric data is often highly sensitive, and a leak of this data can lead to serious privacy breaches. Some of the most sensitive of this type of data relates to the usage of DNA data on individuals. A leak of this type of data without consent could lead to privacy breaches of data protection laws. Along with this, there have been several recent data breaches related to the leak of DNA information, including from 23andMe and Ancestry. It is thus fundamental that a citizen should have the right to know if their DNA data is contained within a DNA database and ask for it to be removed if they are concerned about its usage. This paper outlines a method of hashing the core information contained within the data stores - known as Single-Nucleotide Polymorphisms (SNPs) - into a bilinear group accumulator in batch mode, which can then be searched by a trusted entity for matches.  The time to create the witness proof and to verify were measured at 0.86~ms and 10.90~ms, respectively.}


\onecolumn \maketitle \normalsize \setcounter{footnote}{0} \vfill

\section{\uppercase{Introduction}}
Biometric information often is sensitive as it can reveal personally identifiable information about a person. This could include the recognition of a face, a finger-recognition or a retina. Overall, there is a spectrum of sensitivity without these biometrics, with retina scan and fingerprint recognition often requiring high levels of privacy, whereas it is often difficult to protect one's face from being kept private.  However, one of the most sensitive areas of biometric matching relates to the storage and matching of DNA information. With this, there have been several major data breaches related to DCT-GT (Direct Consumer Testing - Genetic Testing), such as with 23andMe and Ancestry \cite{garner2018privacy}. Along with privacy issues, there is a strong business model for companies selling DNA-related data, such as in 2018 when GlaxoSmith Kline purchased the personal data of thousands of customers from 23AndMe for \$300 million \cite{defrancesco2019your}.

Within DNA analysis, Single-Nucleotide Polymorphism (SNP) information can fully identify a person and which must be thus kept securely. If a person has their SNPs stored in a data store, they often have the right to search for the data in a data store. These types of requests, though, must be kept in a privacy-aware way, and where it is possible to search for SNP information without revealing other information on the contents of a data store of SNP information.

Along with this, this matching could be useful within the Health and Life Insurance industry, as it is possible for a citizen to prove that they do not have a genetic carrier that increases their chances of developing a disease, such as Alzheimer's disease, without disclosing their entire DNA dataset.  

\subsection{Background}
DNA (Deoxyribonucleic acid) is contained in the nucleus of cells and defines the genetic information of a person \cite{bbc}. It is a large and complex polymer which has two strands of a double helix, and where, apart from identical twins, each person has a unique DNA structure. With the cell's nucleus, we have chromosomes which are long threads of DNA. These are then made from genes that contain a code related to a sequence of amino acids - these may then be copied and passed onto the next generation. Each gene can have different forms - which are called alleles. For example, we can have an allele for blue eye colour and another allele for brown eye colour. A genotype is then a collection of these alleles that define a phenotype. Overall, a person inherits one chromosome from their mother and another from their father - and where the pair carry the same gene in the same location.

\subsection{Paper contribution}
The main contribution of this paper is the definition of a privacy-aware approach to the hashing of SNP data into a cryptographic accumulator and then in the evaluation of its performance operation for typical implementations. This type of approach aims to overcome privacy breaches of where private health data data for medical research is shared with insurance companies, such as related to  UK Biobank \cite{Shanti_das}, while also providing proof that an SNP data value has been removed from a data store.

\section{\uppercase{Related work}}

Alsaffar et al. \cite{alsaffar2022digital} outline the risks related to the querying, sharing and genomic testing stages and related countermeasures:

\begin{itemize}
    \item Querying genome data. These risks include the aggregation of data, aggregation of statistics, and correlation attacks. Mitigations include differential privacy, range query limits, homomorphic encryption, and privacy-preserving computing \cite{alsaffar2022digital}.
    \item Sharing genome data. These risks include belief propagation attacks \cite{oksuz2021privacy}, multiparty data sharing, inference attacks, likelihood ratio (LHR) tests \cite{von2019re}, and linkage attacks. The mitigations include statistically aggregated data, multiparty data sharing, encryption, statistical results, multiparty secret sharing, distributed computations, and multiparty queries \cite{alsaffar2022digital}.
    \item Direct to consumer testing. These risks include terms of usage and website vulnerabilities. The mitigations include anonymising genome data and best practice guides \cite{alsaffar2022digital}.
\end{itemize}

Naveed et al.  \cite{naveed2015privacy} defined that genome sequencing technology has advanced at a fast pace and is often focused on: associating with traits and certain diseases, identifying individuals (such as in forensics applications); and in revealing family relationships. Shringarpure et al.  \cite{shringarpure2015privacy} found that in a dataset with 65 individuals, it was possible to pinpoint a certain person from just 250 SNPs.

Bonomi et al. \cite{bonomi2020privacy} outline that sharing genomic data will allow for the enhancement of precision medicine and support providing personalized treatments but privacy concerns and data misuse provide barriers to this sharing. The protection of DNA data provides a challenge for research work as it contains some of the most sensitive attributes of a citizen's identity. In the US, this will typically focus on federal status and regulations such as HIPAA (Health Insurance Portability and Accountability Act) and  GINA (Genetic Information Nondiscrimination Act) \cite{clayton2019law}.  

Bloom filters were first conceived by Harald Bloom \cite{bloom1970space} and allow the data to be hashed by a number of hashing methods, and then the resultant hashes are added to a bit array. They have since been used as a method of storing hashed information on a privacy-aware data store. This includes using segment hashes on disks and which can identify contraband files using a fragment of the file \cite{penrose2015fast}. 

Metsted et al. \cite{melsted2011efficient} used them to count k-mers - which are substrings of length $k$ in DNA sequence data.  For a specific SNP query, privacy-preserving filters have been used to create Privacy-Preserving Polygenic Risk Scores \cite{islam2022privacy}. 

Overall, the core disadvantage of using Bloom filters is that they can only give a confidence level on whether a data item is within a data store and that this confidence level varies with the size of the Bloom filter. They can also lead to large bit arrays, and which can be difficult to scale once values have been hashed into the Bloom data store.

The SIGFRIED approach uses homomorphic encryption to compare two DNA samples between two different sites and which compares genotype data \cite{wang2022privacy}. Blatt et al. \cite{blatt2020optimized} analysed Genome-Wide Association Studies (GWAS\footnote{A GWAS typically involves the study of the interaction between various SNPs and a binary phenotype. This is often related to the disease status of a person}) and identified that privacy issues were a significant barrier to the matching process. To overcome the privacy issues, they outlined the usable of the Cheon-Kim-Kim-Song (CKKS) homomorphic encryption method with a  Residue-Number-System (RNS) variation \cite{cheon2019full} in order to process the SNP data. For 1,000 citizens with 131,071 SNPs, the computation time was 10 minutes on a 28-core computing node. The core disadvantage of homomorphic encryption in relation to the processing of SNPs is the computation overhead. Sim et al. \cite{sim2020achieving} used the HEAAN and SEAL libraries to implement homomorphic encryption and found that it took 24.70 minutes to process a dataset with 245 samples and over four covariates and 10,643 SNPs.

Ayday \cite{ayday2017cryptographic} outlines a method of using homomorphic Signatures, where Alice signs her DNA and stores these signatures. These are then used to provide both privacy and consent on the usage of the data. This uses a modified Paillier public key encryption method.

Differential privacy has also been used to protect sensitive SNP data, such as with   \cite{uhlerop2013privacy}. This work was based on the studies by Homer et al. \cite{homer2008resolving}, and where, under certain conditions, it may be possible to identify a person who had a known genotype within a population of DNA systems - using just minor allele frequencies (MAFs). To overcome this, Uhlerop et al. \cite{uhlerop2013privacy} defined a two-stage that satisfies $\epsilon$-differential privacy. In order to protect against Brute-force attacks on DNA data, honey encryption methods have been proposed, such as with GenoGuard  \cite{huang2015genoguard}. With honey encryption, we can use FPE (Format Preserving Encryption) to hide the original data with the use of an encryption key but then make the ciphertext look like valid SNP data. Thus, when an adversary picks the wrong encryption key, it can reveal valid-looking data but which is incorrect \cite{juels2014honey}.

\section{\uppercase{Accumulators}}

A cryptographic accumulator, as originally proposed by \cite{benaloh1993one}, is a construction that can accumulate a finite set of values into a single succinct accumulator. Accumulators have the advantageous property of being able to efficiently compute a witness, which verifies the membership of any accumulated value in that accumulator. More formally, given a finite set $X = \{x_1, \ldots, x_n \}$,  $\textsf{acc}_X$ is an accumulator of $X$ if every $x \in X$ has an efficiently computable witness $\textsf{wit}_x$  which certifies membership of $x \in X$ and it is computationally infeasible to find a $\textsf{wit}_y$ for any non-accumulated value $y \notin X$ \cite{derler2015revisiting}.

Since the original proposal, the basic notion of a cryptographic accumulator has been extended and iterated upon, resulting in some additional properties. Dynamic accumulators are one such extension, where values can be dynamically added and deleted to and from an accumulator \cite{camenisch2002}. Furthermore, additional security properties such as undeniability and indistinguishability were proposed \cite{derler2015revisiting}. The undeniability security property ensures that it should be infeasible for two contradicting witnesses to be computed, certifying that $\textsf{wit}_x \in \textsf{acc}_X$ and $\textsf{wit}_x \notin \textsf{acc}_X$. Informally, the indistinguishability security property specifies that the accumulator $\textsf{acc}_X$ and witnesses $\textsf{wit}_x$ for $x \in X$ leak no information about the accumulated set $X$. An accumulator that satisfies the indistinguishability property allows computation of a witness that certification of a value in zero-knowledge.

 The original application of a cryptographic accumulator was timestamping of records to ensure their existence as specific times \cite{benaloh1993one}. Over time the use of accumulators has expanded to numerous applications, including ring signatures \cite{xu2010ring}, group signatures \cite{tsudik2003accumulating}, encrypted searches \cite{ge2020secure}, revoking anonymous credentials \cite{xu2019enabling} and vector commitments \cite{catalano2013vector}. Furthermore, the indistinguishability property of some accumulator constructions that allows them to prove in zero-knowledge membership or non-membership of values in the accumulator has been used for revocation of group signatures and anonymous credentials \cite{camenisch2002} such as in the Zerocash protocol, now the Zcash cryptocurrency \cite{bensasson2014}.

\subsection{Hidden Order Group Accumulators}

The original scheme of \cite{benaloh1993one} and its refined variant by \cite{baric1997} was heavily based on the RSA cryptosystem \cite{rivest1978}. The security of Benahloah and de Mare's scheme is derived from the strong RSA assumption (s-RSA) \cite{baric1997}, which states that given two primes $p$ and $q$ of bit-length $\ell$ such that $N = pq$ and a uniformly chosen $c \in \mathbb{Z}^\ast_N$ then it holds that for all probabilistic polynomial time adversaries $\mathcal{A}$ that
\[
    \Pr\left[(m, e) \leftarrow \mathcal{A}(c, n)\ \text{s.t.}\ m^e = c \bmod n \right] \leq \frac{1}{2^\ell}.
\]

Using the s-RSA assumption, the RSA accumulator of \cite{benaloh1993one} and \cite{baric1997} is defined as follows. Let the function $H : \{0, 1\}^\ast \rightarrow \mathbb{Z}^\ast_N$ be a collision-resistant hash function mapping bit strings of arbitrary length to primes in the hidden order set $\mathbb{Z}^\ast_N$. Given a generator $g$ of $\mathbb{Z}^\ast_N$, the RSA accumulator for the set $X = \{x_1, \ldots, x_n\}$ is \[
    \textsf{acc}_A = g^{\prod_{x \in X} H(x)} \bmod N.
\]
A proof of membership that $x_i \in X$ can be created by computing the witness that $\textsf{wit}_{x_i} \in \textsf{acc}_X$ using the following
\[
    \textsf{wit}_{x_i} = g^{\prod_{x \in X\setminus\{x_i\}} H(x)} \bmod N.
\]
Note that the witness is just the RSA accumulator of the set $X\setminus\{x_i\}$ or the $H(x_i)$\textsuperscript{th} root of $\textsf{acc}_X$. Therefore, verification of the witness is done by checking that the equality $\textsf{acc}_X = \textsf{wit}_{x_i}^{H(x_i)} \bmod N$ is satisfied.

The base RSA accumulator was later extended by \cite{camenisch2002} to support dynamically deleting and adding values to and from the accumulator, which became the first known dynamic accumulator construction. Their scheme supported updates of existing witnesses without the required knowledge of the RSA trapdoor. In the work of \cite{li2007}, further improvements were made to the construction by enabling efficient computation of witnesses that function as non-membership proofs. In an exciting improvement, \cite{wang2007} constructed an RSA accumulator that enabled batching of various operations; however, this was shown to be insecure by \cite{camacho2010}.

\subsection{Hash-based Accumulators}

Another common way to construct accumulators is by using symmetric primitives that satisfy the definitions of a collision-resistant hash function, \emph{i.e.} a hash function $H : \{0, 1\}^\ast \rightarrow \{0, 1\}^n$ defined for $n \in \mathbb{N}$ is collision-resistant if for all efficient probabilistic polynomial time adversaries $\mathcal{A}$
\[
    \Pr\left[ (x, y) \leftarrow \mathcal{A}\ \text{s.t.}\ x \neq y\ \text{and}\ H(x) = H(y)\right] \leq \frac{1}{2^n}.
\]
The first constructions that used collision-resistant hash functions were based on hash trees or Merkle trees \cite{merkle1989certified} such as the work of Buldas \emph{et. al.} \cite{buldas2000} who aimed to solve the problems of accountable certificate management. In essence, the hash-based accumulator constructions use a Merkle root to prove the membership of a value within the accumulator \cite{baldimtsi2017accumulators}.

In more recent work by \cite{camacho2008}, the authors developed an accumulator scheme relying only on the collision-resistant hash function assumption (a far weaker assumption than s-RSA). The construction itself uses a hash tree similar to those proposed before it and allows for additions and deletions of values to and from the accumulator. However, despite the greater security guarantees of the collision-resistant hand function assumption, the hash tree accumulator proposed by \cite{camacho2008} was significantly less efficient than the RSA accumulators discussed previously. More recent work has been done to improve the performance of addition and removal operations \cite{reyzin2016}; however, as shown in \cite{camacho2010}, batching is not possible on a number of accumulator operations.

\section{\uppercase{Known Order Group Accumulators With Operation Batching}}

A dynamic accumulator construction was proposed by \cite{nguyen2005accumulators} based on bilinear maps between groups of prime order. Using the notation of \cite{boneh2001} A bilinear map or pairing operation is the map $e : \mathbb{G}_1 \times \mathbb{G}_2 \rightarrow \mathbb{G}_T$ in which $\mathbb{G}_1$, $\mathbb{G}_2$, and $\mathbb{G}_T$ are cyclic groups of prime order $p$ if given the group generators $g \in \mathbb{G}_1$ and $h \in \mathbb{G}_2$ for all $a, b \in \mathbb{F}^\ast_p$ where $\mathbb{F}^\ast_p$ is a finite field over prime $p$ the map $e$ satisfies: 
\begin{itemize}
    \item $e(g^a, h^b) = e(g, h)^{ab} = e(g^b, g^a)$ \hfill (bilinarity)
    \item $e(g, h)$ generates $\mathbb{G}_T$ \hfill (non-degeneracy) 
\end{itemize}
Note that the above is written as for multiplicative groups but can be trivially modified to work with additive cyclic groups of prime order. 

The security of Nguyen's accumulator is based on the $q$-Strong Diffie Hellman ($q$-SDH) assumption. The $q$-SDH assumption, as defined by \cite{boneh2004}, states that given a prime $p$ of bit-length $\ell$, a finite cyclic group $\mathbb{G}$ of order $p$, a generator element $g \in \mathbb{G}$, a uniformly chosen $x \in \mathbb{F}^\ast_p$ and a $q \in \mathbb{N}$ then for all probabilistic polynomial time adversaries $\mathcal{A}$
\[
    \Pr\left[(c, g^{\frac{1}{x + c}}) \leftarrow \mathcal{A}(g, g^x, \ldots, g^{x^q})\right] \leq \frac{1}{2^\ell}
\]
where $c \in \mathbb{F}^\ast_p$ is a scalar of $\mathcal{A}$'s choosing.

The bilinear accumulator, as defined by \cite{nguyen2005accumulators}, $\textsf{acc}_X$ for the finite set $X = \{x_1, \ldots, x_n\}$ is computed as follows:

\begin{mdframed}[linewidth=1pt]
\noindent \underline{$\textsf{acc}.\textsf{setup}$}\\[.5mm]
\noindent Set the public parameters of the bilinear map operation $p$, $\mathbb{G}_1$, $\mathbb{G}_2$, $\mathbb{G}_T$, $e$, $g$, $h$ and uniformly choose a back door $s \in \mathbb{Z}^\ast_p$ and compute $h^s$ for use in the verification process.
\vspace{.5mm}

\noindent \underline{$\textsf{acc}.\textsf{commit}(X)$}\\[.5mm]
\noindent 
Define $H : \{0, 1\}^\ast \rightarrow \mathbb{F}^\ast_p$ to be a collision-resistant hash functions that maps arbitrary length binary strings to elements in the finite field $\mathbb{F}^\ast_p$. The accumulator for the set $X$ is computed by taking the generator for $\mathbb{G}_1$ and computing
\[
    \textsf{acc}_X = g^{\prod_{x \in X}(s + H(x))}
\]
which is essentially computing an accumulator polynomial in the exponent of the $q$-SDH parameters.
\vspace{.5mm}

\noindent \underline{$\textsf{acc}.\textsf{prove}(X, x_ia)$}\\[.5mm]
\noindent The proof of membership witness for the element $x_i \in X$ in the accumulator is computed as
\[
    \textsf{wit}_{x_i} = g^{\prod_{x \in X \setminus \{x_i\}}(s + H(x))}
\]
which is a commitment to the same accumulator polynomial but without the root at $x_i$.
\vspace{.5mm}

\noindent \underline{$\textsf{acc}.\textsf{verify}(\textsf{acc}_X, x_i, \textsf{wit}_{x_i})$}\\[.5mm]
Verification that the element $x_i \in X$ is done by checking that the following pairing operation equailty holds
\[
    e(\textsf{wit}_{x_i}, h^{x_i}h^s) = e(\textsf{acc}_X, h).
\]
\end{mdframed}

\subsection{Addition and Deletion Operations}

In subsequent work by \cite{damgard2008} and \cite{au2009}, the authors expanded the functionality of the \cite{nguyen2005accumulators}, enabling dynamic operations such as the addition and deletion of elements to and from the accumulator as well as non-membership proofs. For use in the rest of the rest of this paper, the addition and deletion operations are defined below.

\begin{mdframed}[linewidth=1pt]
\noindent \underline{$\textsf{acc}.\textsf{add}(\textsf{acc}_X, Y)$}\\[.5mm]
\noindent The set of elements $X \cap Y = \emptyset$ can be added to the accumulator by simply updating the accumulator through the computation
\[
    \textsf{acc}'_X = \textsf{acc}_X^{\prod_{y \in Y}(s + H(y))} = g^{\prod_{z \in X \cup Y}(s + H(z))}
\]

\noindent \underline{$\textsf{acc}.\textsf{delete}(\textsf{acc}_X, Y)$}\\[.5mm]
\noindent The set of elements $Y \subseteq X$ can be deleted from the accumulator by simply updating the accumulator through the computation
\[
    \textsf{acc}'_X = \textsf{acc}_X^{\frac{1}{\prod_{y \in Y}(s + H(y))}} = g^{\prod_{z \in X \setminus Y}(s + H(z))}
\]

\noindent \underline{$\textsf{acc}.\textsf{witnessadd}(\textsf{wit}_{x_i}, x_i)$}\\[.5mm]
\noindent The membership witness $\textsf{wit}_{x_i}$ can be updated on the addition of set $Y$ to correspond to $\textsf{acc}_{X \cup Y}$ by computing
\[
    \textsf{wit}'_{x_i} = \textsf{wit}_{x_i}^{\prod_{y \in Y}(s + H(y))}.
\]

\noindent \underline{$\textsf{acc}.\textsf{witnessdelete}(\textsf{wit}_{x_i}, x_i)$}\\[.5mm]
\noindent The membership witness $\textsf{wit}_{x_i}$ can be updated on the deletion of set $Y$ to correspond to $\textsf{acc}_{x \setminus Y}$ by computing
\[
    \textsf{wit}'_{x_i} = \textsf{wit}_{x_i}^{\frac{1}{\prod_{y \in Y}(s + H(y))}}.
\]
\end{mdframed}

\subsection{Operation Batching}

The result of \cite{camacho2010} shows that batch witness updates with update data of size independent from the number of elements involved is impossible. In many cases, witness updates are important since witnesses that are already issued would become invalid when the accumulator is updated. In work by \cite{vitto2022dynamic} the authors define support for batched operations circumnavigating the impossibility result of \cite{camacho2010} and providing an optimal batch witness update protocol. The batched operations improve the efficient of the dynamic operations of an accumulator for large updates.

\section{\uppercase{SNPs}}
Table \ref{tab:snps} outlines the format of the SNPs \cite{jade_cheng}, and which includes:

\begin{itemize}
    \item RSID field. This field is a unique identifier related to the record. It is not possible to have the same RSID in a raw genome file. Overall, identifiers that start with an 'rs' relate to the dbSNP database sourced from the National Center for Biotechnology Information \cite{wheeler2001database}.
    \item Chromosome. This field can range from 1 to 22 and can also contain coding for X, Y, XY, and MT.
    \item Position. This field typically ranges from 3 to 249,218,992, and where there can be multiple matches for the same chromosome and position fields.
    \item Genotype. This field relates to the genotype of the SNP.
\end{itemize}

\begin{table}[ht]
\centering
\caption{SNP data format \cite{jade_cheng}}
\label{tab:snps}
\begin{tabular}{l | l | l | l }
\hline
RSID  &      Chromosome &  Position &  Genotype\\
\hline\hline
rs12564807 &   1       &      734462   &  AA\\
rs3131972  &   1       &     752721   &  AG\\
rs148828841 &  1       &     760998   &  CC\\
rs12124819  &  1       &     776546   &  AA\\
rs115093905 &  1       &     787173   &  GG \\  
\hline
\end{tabular}
\end{table}

For instance, with the SNP record of "rs367789441 1 68082 TT", the private element of this data is the genotype "TT", and where the two bases ("TT")  come from a person's "X" and "Y" allele.

Overall, the main matches that relate to these searches often relate to genetic relatedness, kinship, parentship and medical ailments. For example, "rsID rs429358" is related to the APOE-$\epsilon$4 allele and which has a strong influence on the risk of Alzheimer's disease \cite{lundberg2023novel}.

\section{\uppercase{Privacy-aware searching}}
The number of SNPs stored within a data store can vary based on the applications. A \emph{personalized hair treatment} use case can range between 30 to 40 SNPs, and ancestry applications require around 350,000 SNPs. Overall, a person will have around 650,000 SNPs. 

Overall, Company X could hash all of its SNPs, such as for "rs12564807-1-734462-AA", and store them in an accumulator and pass this to Trent.  A basic use case could be where a person (Alice) wants to know if Company X has their SNPs on their data store. Company X then fills a privacy-aware store with the SNPs they have and gives it to a trusted entity (Trent) - each SNP is then hashed into the accumulator. Alice sends hashed versions of the SNPs and asks Trent to search for them. Trent then returns proof that all the SNPs are contained in the store or not. This can be 25 matches for a paternity test and 44 for a person. The use of an accumulator, too, could provide a way of showing Alice that her SNPs have been removed from Company X's datastore.

\begin{figure*}[ht]
\begin{center}
\includegraphics[width=.9\linewidth]{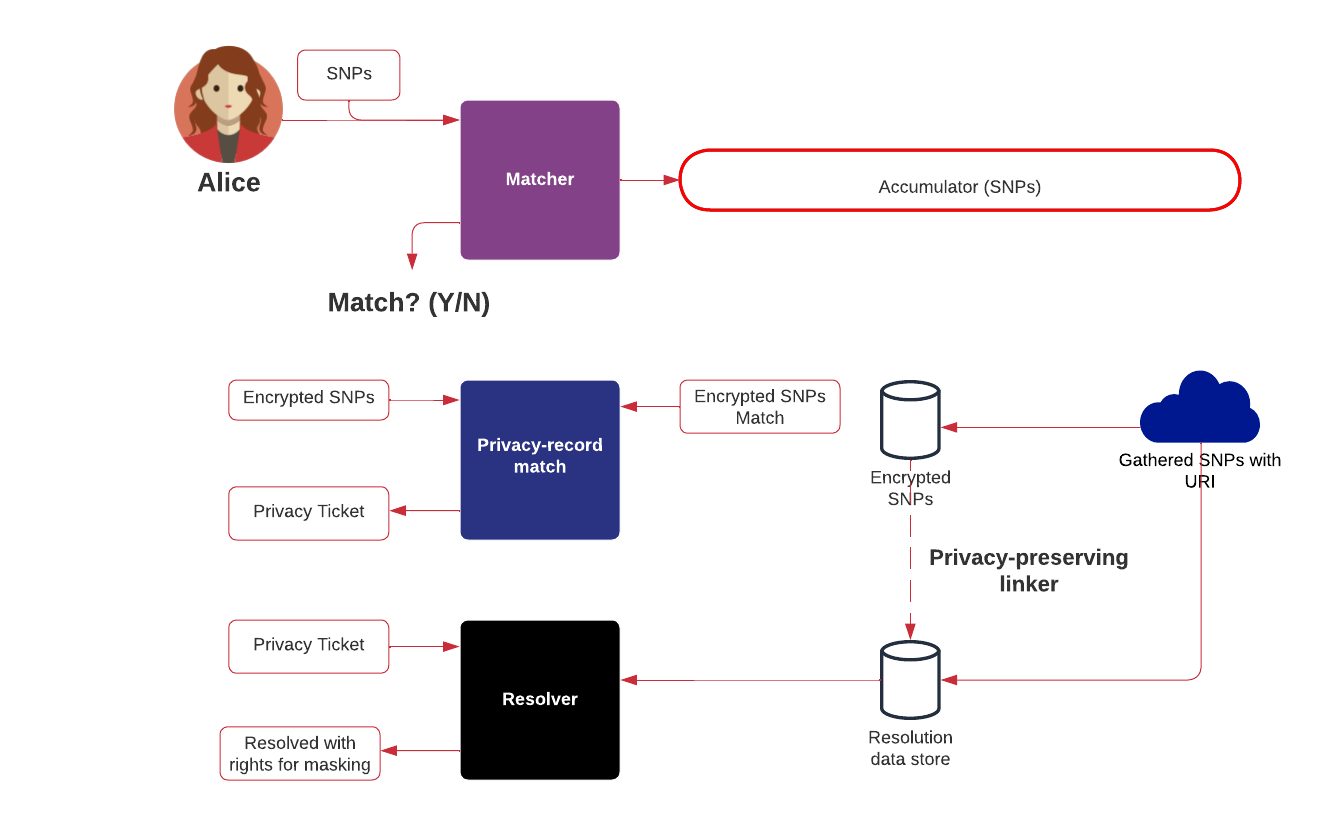}
\caption{Privacy-aware framework for SNP matching}
\label{fig:snps}
\end{center}
\end{figure*}

Figure \ref{fig:snps} outlines the privacy-aware framework. The system splits into two main systems: Matcher; and Resolver, and has three main stages:

\begin{itemize}
\item Setup:  Initially, the gathered SNPs are encrypted and stored with a unique identifier. This identifier is then passed to a resolution service and which stores the details related to the gathered data for the SNPs. Each gathered set of SNPs is split either into hash values or split into secret shares and then added into an accumulator.
\item Matching: When Alice wants to see if her SNPs have been registered on the system, she submits these and is split into hash values or shares. The Bloom filter will then say for definite if SNPs are not stored. With a high probability of success, Alice will be informed if there is a successful discovery. Alice’s encrypted set of SNPs is then submitted to be checked against a full set of the previously encrypted SNPs. A hash search will find the matches.
 \item Resolving: On one or more matches, Alice will be delivered a privacy ticket which will be used to resolve the actual details of SNPs gathering within the resolver service.
\end{itemize}

\section{\uppercase{Results}}
The RSA accumulator method would struggle with performance in adding and removing so many hashed values of SNPs, but the pairing-based method using the BLS 12381 curve \cite{barreto2003constructing} provides a more efficient way. For this, we hash our data onto the curve and produce a scalar value which can then be added into the accumulator \cite{asecuritysite_96654}. Table \ref{tab:snps2} shows the results of using the method defined in \cite{vitto2022dynamic} and running on a t2.medium instance in AWS (two vCPUs and 4GB of memory) for the operations of hashing to the curve and adding to the accumulator, and also where we have pre-computed hashes and then just add these to the accumulator. In this case, one hash is added to the accumulator at a time. We see, though, that the addition of the hashing of the SNP onto the curve has a minimum effect compared with the adding of the hashed value into the accumulator. For this, the experiment gives 68.23 seconds for the addition of 100,000 SNPs.

In Table \ref{tab:snps3}, we use the batch mode of the Vitto et al. method \cite{vitto2022dynamic}, and where we can add in batches of 10 and 100. We can see that batch processing considerably enhances the speed of operation of adding the SNP hashes to the accumulator. `Now, rather than 68.23 seconds for the addition of 100,000 SNPs, the time to compute drops to 0.87 seconds.

Overall, it can be seen that for the range of examples, there is an almost linear relationship in adding the SNP into the accumulator, and has an approximate overhead of around 0.774~ms for adding a single SNP hash to the accumulator, and 8.7~$\mu$S when processed in batches of 100. For batches of 100, it would mean that a single user with 640,000 SNPs onto the accumulator - in this case - would take around 5.65~seconds. The time to create the witness proof and to verify does not vary as much as the number of values in the accumulator, and these were measured at 0.86~ms and 10.90~ms, respectively. 

\begin{table*}[ht]
\centering
\caption{Results for adding to accumulator }
\label{tab:snps2}
\begin{tabular}{l | l | l | l | l}
\hline
Method  &    Adding 100 (ms)	& Adding 1,000	(ms) & Adding 10,000 (ms)	& Adding 100,000 (ms)\\
\hline\hline
Adding with hash &	0.1	& 0.9	& 7.28 &	69.44\\
Adding no hash	& 0.1	& 0.87	& 7.01	& 68.23\\
\hline
\end{tabular}
\end{table*}

\begin{table*}[ht]
\centering
\caption{Results for adding to accumulator using batch mode}
\label{tab:snps3}
\begin{tabular}{l | l | l | l | l}
\hline
Method  &    Adding 100 (ms)	& Adding 1,000	(ms) & Adding 10,000 (ms)	& Adding 100,000 (ms)\\
\hline\hline
Batch of 1	& 0.1 & 	0.66 &  6.95 & 67.3\\
Batch of 10	& 0.01	&  0.1	&  0.86	&  7.61\\
Batch of 100 & 0.006 & 0.01	 & 0.12	&  0.87\\
\hline
\end{tabular}
\end{table*}

\section{\uppercase{Conclusions}}
While Bloom filters and homomorphic encryption provide strong levels of privacy, they can be affected by scalability issues. The usage of accumulators can provide one method of preserving the data values in a data store, and for this, to have a fixed data size width. The hashing of the SNP onto an elliptic curve point on the BLS 12381 curve has a relatively small overhead when compared with the addition of the point into the accumulator. The time taken to build the accumulator is thus mostly dependent on the time to add the data point into the accumulator. The batch mode used in \cite{ayday2017cryptographic} considerably reduces the processing overhead when adding the SNPs when used in batches. Overall, one of the core advantages of using an accumulator is that we can provide witness proof that a data entity is contained within a data set, along with the entity being removed. 

\bibliographystyle{apalike}
\bibliography{references}

\end{document}